\documentstyle[12pt]{article}
\newcommand{\doublespace}{\renewcommand{\baselinestretch}{1.75}
   \Large\normalsize}

\renewcommand{\ref}[1]{\raisebox{.6ex}{[#1]}}

\newcommand{\be}{\begin{equation}}
\newcommand{\ee}{\end{equation}}

\begin{document}

\doublespace

\title{ Nonlinear Schr\"odinger Equation for Superconductors }

\author{Ping Ao, David J. Thouless, and X.-M. Zhu    \\
Department of Physics, FM-15                 \\
University of Washington, Seattle, WA 98195  \\ }

\maketitle

\begin{abstract}
Using the Hartree-Fock-Bogoliubov factorization of the density matrix
and the Born-Oppenheimer approximation
we show that the motion of the condensate satisfies a
nonlinear Schr\"odinger equation in the zero temperature limit.
The Galilean invariance of the equation is explicitly manifested.
{}From this equation some general properties of a superconductor,
such as Josephson effects, the Magnus force,
and the Bogoliubov-Anderson mode can be obtained readily.
\end{abstract}

\noindent
PACS${\#}$s: 74.20.-z; 67.50.Fi;  \\

\newpage

A large class of superconducting phenomena is well described by a static
equation near superconducting transition temperature,
proposed by Ginzburg and
Landau (GL)\cite{gl} before the microscopic theory of Bardeen, Cooper,
and Schrieffer\cite{bcs}. Subsequently the GL equation has been derived
microscopically by Gorkov\cite{gorkov} and extended to a wider parameter region
by many authors\cite{werthamer,cyrot}.
Despite the tremendous success of the microscopic theory,
the GL equation occupies an important position,
as also illustrated in the recent study
of the clarifying of symmetry of the Cooper
paring in high Tc superconductors\cite{annett},
of the flux phases\cite{huse}, and of the insulator-superconductor
transitions\cite{fisher}.
Encouraged by the wide region of validity of the GL equation,
several attempts to derive the equation of motion for the condensate
in a superconductor, the so-called the time-dependent GL equation,
have been made
during the past three decades\cite{werthamer,cyrot,zwerger,stoof}.
Although there is no consensus for the exact form,
a concern about the apparent non-Galilean invariance
in the previous derived equations of motion exists as noted
in Refs.\cite{werthamer,stoof}.
While it can be justified near the transition temperature because of the strong
relaxation, it is questionable in the low temperature limit.
On the other hand,
the Galilean invariant form of the equation of motion has been repeatedly
argued
phenomenologically:
the proposal of Ref.\cite{feynman} in the derivation of Josephson relations;
the proposal of Ref.\cite{fisher}
in the study of insulator-superconductor transitions, and the proposals based
on symmetry considerations\cite{greiter,peng}.
Besides the conceptual question of the Galilean invariance, the different
forms of the equation of motion have different physical consequences,
as shown, for example, in the study
of the decay of supercurrent\cite{ao} and of the Hall effect\cite{dorsey}.
The requirement of the Galilean invariance will put a restriction on the form
of the equation of motion.
Furthermore, it guarantees certain global
properties of a superconductor such as Josephson effects and the existence of
the Magnus force.

The purpose of the present paper
is to give a formal microscopic derivation of
the equation of motion for the condensate near zero temperature,
which explicitly manifests Galilean invariance.
The present derivation is based on the existence
of the off-diagonal long range order(ODLRO)
in a superconductor\cite{yang} which allows a macroscopic wavefunction
description of the condensate, and on the Born-Oppenheimer approximation which
allows the separation of the slow motion of the condensate and the fast
motion of the formation of the Cooper pairs.
A nonlinear Schr\"odinger equation for the condensate is obtained.
The usefulness of this equation is illustrated by three examples:
Josephson effects, the Magnus force, and the Bogoliubov-Anderson mode.
For definiteness, a clean s-wave superconductor will be considered.

To demonstrate the generality of the present derivation and to give a firm
microscopic ground, we first derive a set of self-consistent equations for the
one-body reduced density matrix and the wavefunction describing ODLRO.
The model Hamiltonian for $N$ electrons may be taken as
\be
   H = \sum_{j=1}^{N} \frac{ - \hbar^{2} }{2m}
       \nabla_{j}^{2}
       + \frac{1}{2} \sum_{j\neq l}^{N} U({\bf r}_{j}-{\bf r}_{l}) \; .
\ee
Here $m$ is the mass of an electron,
${\bf r} = (x,y,z)$, and $\nabla$ is the three-dimensional gradient.
The effect of the lattice is to renormalize the electron mass and to give
rise a weak, short range, and attractive two body interaction
$U$ with no spin dependence, which is responsible for superconductivity.
Equation (1) is in the standard form of the Hamiltonian in the derivation of
BCS theory.
Here we ignore the effect of the electromagnetic field, and
we shall return to it later.
The motion of electrons is described by the $N$-body density matrix
$D({\bf r}_{1}\sigma_{1}, ... , {\bf r}_{N}\sigma_{N};
{\bf r}_{1}'\sigma_{1}', ... , {\bf r}_{N}'\sigma_{N}'; t ) $,
which satisfies the von Neumann equation
\be
   i\hbar \frac{\partial}{\partial t} D    = [ H, D] \; .
\ee
Here $\sigma_{j} = \pm$ is the spin index of $j$th electron.
The $K$-body reduced density matrix is defined as
\[
   R_{K}({\bf r}_{1}\sigma_{1}, ... , {\bf r}_{K}\sigma_{K};
     {\bf r}_{1}'\sigma_{1}', ... , {\bf r}_{K}'\sigma_{K}'; t )
   = \frac{N!}{(N-K)!}
     \prod_{l=K+1}^{N} \sum_{\sigma_{l}=\pm}\int d^{3}{\bf r}_{l} \times
\]
\be
    D({\bf r}_{1}\sigma_{1}, ... , {\bf r}_{N}\sigma_{N};
     {\bf r}_{1}'\sigma_{1}', ... , {\bf r}_{K}'\sigma_{K}',
     {\bf r}_{K+1}\sigma_{K+1}, ... , {\bf r}_{N}\sigma_{N}; t ) \; .
\ee
The information needed is contained in the one- and two-body reduced density
matrices.
Then, from eqs. (2) and (3), we have the equations of motion
for the one-body reduced density matrix $R_{1}$,
\[
   i\hbar \frac{\partial}{\partial t}
   R_{1}({\bf r}_{1}\pm; {\bf r}_{1}'\pm; t )
    =  - \frac{\hbar^{2} }{2m} [ \nabla_{1}^{2} -  \nabla_{1}'^{2} ]
      R_{1}({\bf r}_{1}\pm; {\bf r}_{1}'\pm; t )
\]
\be
   + \int d^{3}{\bf r}_{2} \sum_{\sigma_{2}= \pm }
   [ U({\bf r}_{1} - {\bf r}_{2} )
   - U({\bf r}_{1}' - {\bf r}_{2} ) ] R_{2}({\bf r}_{1}\pm,
       {\bf r}_{2}\sigma_{2}; {\bf r}_{1}'\pm, {\bf r}_{2}\sigma_{2}; t ) \; .
\ee

Now, we assume that two-body and higher density matrices can be decomposed
in terms of one-body reduced density matrix with ODLRO. This is
the usual generalization of Dirac's\cite{dirac}
factorization which leads to the time-dependent Hartree-Fock theory.
Specifically, we have for the triplet part of the two-body reduced density
matrix $R_{2}$,
\be
  R_{2}({\bf r}_{1}\pm, {\bf r}_{2}\pm ;
                 {\bf r}_{1}'\pm, {\bf r}_{2}'\pm; t )
  \approx   R_{1}({\bf r}_{1}\pm; {\bf r}_{1}'\pm; t )
            R_{1}({\bf r}_{2}\pm; {\bf r}_{2}'\pm; t )
         -  R_{1}({\bf r}_{1}\pm; {\bf r}_{2}'\pm; t )
            R_{1}({\bf r}_{2}\pm; {\bf r}_{1}'\pm; t ) \; .
\ee
Because of the existence of the existence of ODLRO, for the singlet part we
have
\be
   R_{2}({\bf r}_{1}\pm, {\bf r}_{2}\mp ;
                 {\bf r}_{1}'\pm, {\bf r}_{2}'\mp; t )
   \approx  R_{1}({\bf r}_{1}\pm; {\bf r}_{1}'\pm; t )
            R_{1}({\bf r}_{2}\mp; {\bf r}_{2}'\mp; t )
   +      \Psi({\bf r}_{1}\pm, {\bf r}_{2}\mp; t )
          \Psi^{\ast}({\bf r}_{1}'\pm, {\bf r}_{2}'\mp; t ) \; .
\ee
The existence of the wavefunction $\Psi$ is the manifestation of ODLRO.
Here, as $|{\bf r}-{\bf r}'| \rightarrow \infty$,
$R_{1}({\bf r}_{1}\pm; {\bf r}_{1}'\pm; t ) \rightarrow 0$ exponentially.
In the absence of spin interaction, we have
$R_{1}({\bf r}_{1}\pm; {\bf r}_{1}'\pm; t )
     = R_{1}({\bf r}_{1}; {\bf r}_{1}'; t )$,
independent of spin index, and $R_{1}({\bf r}_{1}\pm; {\bf r}_{1}'\mp; t ) =
0$.
Therefore, from eqs. (4)-(6) we have for the one-body reduced density
matrix $R_{1}$
\[
   i\hbar \frac{\partial}{\partial t}
   R_{1}({\bf r}_{1}; {\bf r}_{1}'; t )
    \approx - \frac{\hbar^{2} }{2m} [ \nabla_{1}^{2} - \nabla_{1}'^{2} ]
                     R_{1}({\bf r}_{1}; {\bf r}_{1}'; t )
\]
\[
   + \int d^{3}{\bf r}_{2} [ U({\bf r}_{1} - {\bf r}_{2} )
   - U({\bf r}_{1}' - {\bf r}_{2} ) ] 2 R_{1}({\bf r}_{2};{\bf r}_{2}; t )
     R_{1}({\bf r}_{1};{\bf r}_{1}'; t )
\]
\[
   - \int d^{3}{\bf r}_{2} [ U({\bf r}_{1} - {\bf r}_{2} )
   - U({\bf r}_{1}' - {\bf r}_{2} ) ] R_{1}({\bf r}_{1};{\bf r}_{2}; t )
     R_{1}({\bf r}_{2}; {\bf r}_{1}'; t )
\]
\be
   + \int d^{3}{\bf r}_{2} [ U({\bf r}_{1} - {\bf r}_{2} )
   - U({\bf r}_{1}' - {\bf r}_{2} ) ] \Psi({\bf r}_{1}, {\bf r}_{2}; t )
       \Psi^{\ast} ({\bf r}_{1}', {\bf r}_{2}; t ) \; .
\ee
A similar factorization of the equation of motion for $\Psi$ gives
\[
   i\hbar \frac{\partial}{\partial t}
   \Psi({\bf r}_{1}, {\bf r}_{2}; t )
   \approx - \frac{\hbar^{2} }{2m} [ \nabla_{1}^{2} + \nabla_{2}^{2}]
   \Psi({\bf r}_{1}, {\bf r}_{2}; t )
\]
\[
     + \left[ U({\bf r}_{1} - {\bf r}_{2} )
         + \int d^{3}{\bf r}_{3} [ U({\bf r}_{1} - {\bf r}_{3} )
                                   + U({\bf r}_{2} - {\bf r}_{3} )
       2 R_{1}({\bf r}_{3}; {\bf r}_{3}; t ) ] \right]
         \Psi( {\bf r}_{2}',{\bf r}_{3}; t )
\]
\be
   + \int d^{3}{\bf r}_{3}
     [ U({\bf r}_{1} - {\bf r}_{3} ) + U({\bf r}_{2} - {\bf r}_{3} ) ]
     [ R_{1}({\bf r}_{1}; {\bf r}_{3}; t )
        \Psi({\bf r}_{2}, {\bf r}_{3}; t )
     + R_{1}({\bf r}_{2}; {\bf r}_{3}; t )
        \Psi({\bf r}_{1}, {\bf r}_{3}; t ) ]\;.
\ee
Here $ \Psi({\bf r}_{1}, {\bf r}_{2}; t )
 = [ \Psi({\bf r}_{1}+, {\bf r}_{2}-; t )-
     \Psi({\bf r}_{1}-, {\bf r}_{2}+; t ) ]/\sqrt{2}$, and is symmetric upon
interchange of ${\bf r}_{1}, {\bf r}_{2}$ because of the singlet paring.

Eqs. (7) and (8) form a general self-consistent set of equations valid at any
temperature.
We note that eqs. (7) and (8) are similar to the Gorkov equations in the form
given by Keldysh Green functions\cite{volkov},
but here we use reduced density matrices,
following the approach of Penrose\cite{penrose} in the case
of superfluid helium 4.
Same equations have been obtained previously\cite{ufg}.
We emphasize that the presentation of leading to eqs. (7) and (8)
is to show the generality of the following nonlinear Schr\"odinger equation
for the condensate.

For an illustration of the Galilean invariance we consider a uniform state.
The uniform solution of eqs. (7) and (8)
has the form for the one-body density matrix $R_{1}$
\be
   R_{1}( {\bf r}_{1};{\bf r}_{1}'; t )
  = \rho ( {\bf r}_{1}-{\bf r}_{1}' ) \; ,
\ee
and for the wavefunction $\Psi$ describing ODLRO
\be
   \Psi( {\bf r}_{1},{\bf r}_{2}; t ) = \Psi_{0}( {\bf r}_{1}-{\bf r}_{2} )
       \exp \left\{ - \frac{ i }{\hbar} V_{0} \; t \right\}  \; ,
\ee
which depends only on the relative coordinate\cite{taylor}.
Then it is easy to check that the uniform current carrying state,
${\bf v} = \hbar{\bf k} /m $, is also a solution of eqs.(7) and (8).
In this case,
\be
   R_{1}({\bf r}_{1};{\bf r}_{1}'; t ) = \rho({\bf r}_{1}-{\bf r}_{1}')
     \exp\{ i {\bf k} \cdot ({\bf r}_{1} - {\bf r}_{1}' ) \} \; ,
\ee
and
\be
   \Psi( {\bf r}_{1},{\bf r}_{2}; t ) = \Psi_{0}( {\bf r}_{1}-{\bf r}_{2} )
     \exp \left\{ - \frac{ i }{\hbar} V_{0} \; t \right\}
     \exp \left\{ i {\bf k} \cdot ({\bf r}_{1} + {\bf r}_{2} )
        - i \frac{\hbar {\bf k}^{2} }{m} t \right\} \; ,
\ee
and eqs. (7) and (8) remain unchanged.

To derive the equation of motion for the condensate,
the theme of the present paper,
we observe that eqs. (7) and (8) describe three kinds of distinct dynamics:
the motion of the quasiparticle excitations,
the formation of Cooper pairs, and the motion of the condensate.
At low enough temperatures, because of the existence of the energy gap in the
quasiparticle energy spectrum, the density of quasiparticle excitations
is exponentially small.
In this case, we only need to deal with the dynamics of the formation of
Cooper pairs and the motion of Cooper pairs.
To avoid unnecessary complications, we shall work under this restriction.

As suggested by the demonstration from eqs.(9) and
(10) to eqs. (11) and (12), we observe that
the formation of a Cooper pair is described by the relative coordinate
${\bf r}={\bf r}_{1} - {\bf r}_{2}$.
The motion of the condensate is described by the
center of mass coordinate ${\bf R} = ({\bf r}_{1} + {\bf r}_{2} )/2$.
If the motion of the condensate is slow enough, which in principle can be
arbitrarily slow,
a separation of dynamics between the motion of the condensate
and the formation of a Cooper pair is possible.
The Born-Oppenheimer approximation can then
be applied to get a simple equation of motion for the condensate,
because the dynamics of the formation of a Cooper pair
can always follow the motion of the condensate.
To implement the Born-Oppenheimer approximation\cite{schiff},
the wavefunction $\Psi$ may be written as
\be
   \Psi({\bf r}_{1}, {\bf r}_{2}; t ) = \psi({\bf R}; t ) \;
                      \phi({\bf r}, {\bf R}, \psi,\psi^{\ast}; t ) \; ,
\ee
with the requirement that
\be
   \int d^{3}{\bf r}\; |\phi({\bf r},{\bf R},\psi,\psi^{\ast}; t )|^{2} = 1\; ,
\ee
because the wavefunction $\phi$ for the formation of a Cooper pair
is bounded in space.
With eq. (14), the wavefunction of the condensate $\psi$ is normalized to the
density of Cooper pairs, which is proportional to the superfluid density.
Therefore, from eq. (8)
according to the Born-Oppenheimer approximation\cite{schiff}
the wavefunction $\phi$ satisfies following stationary equation
\[
   V({\bf R}, \psi, \psi^{\ast} ) \; \phi({\bf r}) =
    -\frac{\hbar^{2}}{m} \nabla_{\bf r}^{2}
                    \phi({\bf r}) +  U({\bf r})\phi({\bf r})
\]
\[
  + \int d^{3}{\bf r}_{3} 2 R_{1}({\bf r}_{3};{\bf r}_{3} )
    \left[ U({\bf R}+\frac{ {\bf r} }{2} - {\bf r}_{3})
         + U({\bf R}-\frac{ {\bf r} }{2} - {\bf r}_{3}) \right] \phi({\bf r})
\]
\[
   + \int d^{3}{\bf r}_{3} 2 R_{1}({\bf r}_{3};{\bf r}_{3} )
     \left[ U({\bf R}+\frac{ {\bf r} }{2} - {\bf r}_{3})
          + U({\bf R}-\frac{ {\bf r} }{2} - {\bf r}_{3}) \right] \times
\]
\[
       \left[ R_{1}( {\bf R} + \frac{ {\bf r} }{2}; {\bf r}_{3} )
   \frac{\psi( ({\bf R} - \frac{ {\bf r} }{2} + {\bf r}_{3})/2; t ) }
        {\psi({\bf R}; t ) }
    \phi({\bf R}- \frac{ {\bf r} }{2} - {\bf r}_{3} ) \right.
\]
\be
  \left.    - R_{1}( {\bf R} - \frac{{ \bf r} }{2}; {\bf r}_{3} )
   \frac{\psi( ({\bf R} + \frac{ {\bf r} }{2} + {\bf r}_{3})/2; t ) }
        {\psi({\bf R}; t ) }
    \phi({\bf R} + \frac{ {\bf r} }{2} - {\bf r}_{3} ) \right] \; ,
\ee
with $R_{1}$ solved from eq. (7) as a function of $\psi (\psi^{\ast})$ and
$\phi (\phi^{\ast})$.

Given the slow varying quantities ${\bf R}$ and $\psi (\psi^\ast)$,
eqs.(15) and (7) can
be solved in the same manner as the usual Born-Oppenheimer approximation.
In the special case of a uniform state, those equations have been solved
in Ref.\cite{taylor}(see also Ref.\cite{ufg}),
where it is found that $V = 2 \mu$ with $\mu$ the Fermi energy of the system,
and that $\Psi$ determined by the gap equation
is proportional to the electron number density $n$.
In the more general case there will be a small correction to $2\mu$ due to
variation of the Cooper pair binding energy (the energy gap)
as a function of $|\psi|$.
Now suppose that we can solve eq. (15) for $V$ and $\phi$
in terms of ${\bf R}$ and $\psi \; (\psi^{\ast})$ generally.
Using eqs. (8), (13)-(15) we find the wavefunction $\psi$ of the condensate
satisfies the following equation:
\be
  i\hbar \frac{\partial }{\partial t} \psi({\bf R}; t) = \frac{1}{4m}
   [ - i\hbar\nabla_{\bf R} + {\bf a}({\bf R},\psi,\psi^{\ast}) ]^{2}
    \psi({\bf R}; t)
   + [a_{0}({\bf R},\psi,\psi^{\ast})+V({\bf R},\psi,\psi^{\ast})]
    \psi({\bf R}; t) \; ,
\ee
with
\be
   {\bf a}({\bf R},\psi,\psi^{\ast}) = - i\hbar\int d^{3}{\bf r} \;
     \phi^{\ast}({\bf r},{\bf R}, \psi, \psi^{\ast} )
     \nabla_{\bf R} \phi({\bf r},{\bf R}, \psi, \psi^{\ast} ) \; ,
\ee
and
\be
   a_{0}({\bf R},\psi,\psi^{\ast}) = \frac{\hbar^{2} }{4m} \int d^{3}{\bf r} \;
      | \nabla_{\bf R} \phi({\bf r},{\bf R}, \psi, \psi^{\ast}) |^{2}
              - \frac{1}{4m} {\bf a}^{2}({\bf R}, \psi,\psi^{\ast}).
\ee
We note that the mass of a Cooper pair in eq.(16) is twice of the electron
mass, and eq.(16) is similar to  the form in the
discussion of the separation of fast and slow motions\cite{berry}.
The the pseudo-scalar and vector potentials $a_{0}$ and ${\bf a}$
are the response of the fast internal degrees of freedom,
the dynamics of the formation of a Cooper pair,
to the slow motion of the condensate.
Because of the singlet pairing, $\phi$ is spherically symmetric in a uniform
state. In this case both ${\bf a} = 0$ and $a_{0} = 0$.
For a nonuniform state, the shape of a Cooper pair wavefunction $\phi$
is a deformed sphere, and we may choose $\phi$ to be real. In this case
we have ${\bf a}=0$ but $a_{0} \neq 0$.
This is consistent with the generalized GL equation\cite{werthamer}.
We point out that for non singlet pairing the pseudo-vector potential ${\bf a}$
may not be zero.
Equation (16) is our desired equation of motion for the condensate, which takes
the form of a nonlinear Schr\"odinger equation, and is Galilean invariant.

The coupling to the electromagnetic field can be introduced as usual as the
minimum coupling into the Hamiltonian eq. (1).
Performing a parallel calculation,
we find that the equation of motion of the condensate in this case
satisfies the following equation:
\[
  i\hbar \frac{\partial }{\partial t} \psi({\bf R}; t) = \frac{1}{4m}
   [ - i\hbar\nabla_{\bf R} - 2e {\bf A}({\bf R})
     + {\bf a}({\bf R},\psi,\psi^{\ast} ) ]^{2}
           \psi({\bf R}; t)
\]
\be
   + [ 2e A_{0}({\bf R}) + a_{0}({\bf R},\psi,\psi^{\ast})
   + V({\bf R}, \psi,\psi^{\ast} ) ] \psi({\bf R}; t) \; .
\ee
To reach the above equation,
we have assumed in addition that the space variation of the scalar gauge
potential $A_{0}$ and the vector gauge potential ${\bf A}$
should be smooth over the length scale of the
localization length of a Cooper pair to ignore the direct
influence of the gauge potentials on the formation of Cooper pairs.
This condition restricts eq. (19) to the case of
a type II superconductor. Supplemented by Maxwell's equations for the
electromagnetic field, the properties of a superconductor can be discussed
under the above conditions.

Having obtained the nonlinear Schr\"odinger equation for the condensate,
eq.(16) or (19), we consider in the following their consequences,
Josephson effects, the Magnus force, and the Bogoliubov-Anderson mode.
In the first two cases the dynamics of the the condensate
is dominated by the phase of the condensate wavefunction $\psi$.
The last case is concerned with the compressibility of the superconducting
state, involving both the the phase and the amplitude of the condensate
wavefunction $\psi$.

{\it Josephson relations}.
The derivation of Josephson relations in Ref.\cite{feynman}
is the neatest one known so far, which is
started by postulating of a Schr\"odinger equation for the motion of the
condensate. However, no justification for the Schr\"odinger equation
in Ref.\cite{feynman} has been provided. In particular,
one also needs to justify that the electric charge is $2e$ instead of $e$.
The present derivation gives such answers. We note that
by AC Josephson effect and the DC Josephson interference effect, the $2e$ of a
Cooper pair is clearly established.

{\it The existence of the Magnus force}.
Since our equation of motion of the condensate, eq.(16),
is identical to the form of
Gross-Pitaevskii equation for superfluid helium 4,
the existence of the Magnus force for a vortex line in a superconductor,
which is proportional to the number density of the superfluid electrons,
follows naturally\cite{dorsey}.
This  coincides with the result from a complete different
approach\cite{thouless}, therefore gives another general demonstration of the
existence of the Magnus force.
By study of the consequences, such as the Hall effect\cite{dorsey},
of the Magnus force, we may have a direct measurement of the superfluid
velocity therefore
superfluid number density in comparison to the London penetration depth.

{\it The Bogoliubov-Anderson mode}.
A nonlinear Schr\"odinger equation may
contain a phonon-like excitation mode\cite{gross}.
We show below that this mode in eq.(16) has a propagating
velocity of $v_F/\sqrt{3}$ with $v_{F}$ the Fermi velocity in the neutral
superconductor,
and is the plasma mode in the charged case of eq.(19).
We write the condensate wavefunction as $\psi = \sqrt{\rho} \;
\exp\{ \theta \}$.
Ignoring the small correction due to the variation of the binding energy of a
Cooper pair we have $V = 2 \mu(n)$.
Expanding the phase and the Cooper pair density around their uniform state
values, $\theta = \theta_0 + \delta \theta$
and $\rho = \rho_0 + \delta \rho$, Using $\rho \propto n$, and eliminating
$\delta\rho$ in eq.(16), we obtain
\be
    \left[ \frac{3}{ v_{F}^{2} } \frac{\partial^{2} }{\partial t^{2} }
     - \nabla_{\bf R}^{2} \right] \delta\theta ({\bf R}; t) = 0 \; .
\ee
This describes the Bogoliubov-Anderson mode,
as known by Refs.\cite{werthamer,stoof}.
We note that the expansion leading to eq.(20) effectively
chooses a reference frame.
In the derivation of eq.(20), the relationship between the Fermi energy $\mu$
and the electron number density $n$
in the weak interaction limit has been used.
For a real superconductor eq.(19) should be used together with Maxwell's
equations. Because the scale gauge potential $A_0$
depends on the superfluid electron number density,
we recover the usual plasma mode.

In conclusion, we have derived a nonlinear Schr\"odinger equation
for the superconducting condensate near zero temperature,
which is Galilean invariant. The Born-Oppenheimer approximation employed in the
derivation allows a natural separation of the fast formation of a Cooper pair
and the slow motion of the condensate.
By the nonlinear Schr\"odinger equation,
Josephson effects, the existence of the Magnus force,
and the Bogoliubov-Anderson mode in a superconductor can be readily
established. This shows a unified physics behind those phenomena.

\noindent
{\bf Acknowledgements:}
One of us (P.A.) thanks A.J. Leggett for useful suggestions and
L.-Y. Shieh for helpful discussions.
This work was supported in part
by US National Science Foundation under Grant No's. DMR
89-16052 and DMR 92-20733.


\begin{thebibliography}{99}
\bibitem{gl}
 V.L. Ginzburg and L.D. Landau, Zh. Eksp. Teor. Fiz. {\bf 20}, 1064 (1950).
\bibitem{bcs}
 J. Bardeen, L.N. Cooper, and J.R. Schrieffer, Phys. Rev. {\bf 108},
  1175 (1957).
\bibitem{gorkov}
 L.P. Gorkov, Zh. Eksp. Teor. Fiz. {\bf 36}, 1918 (1959).
\bibitem{werthamer}
 See, N.R. Werthamer, in {\it Superconductivity}, Vol. 2,
 edited by R.D. Parks, M. Dekker, New York, 1969, for a review before 1968.
\bibitem{cyrot}
 See, M. Cyrot, Rep. Prog. Phys. {\bf 36}, 103 (1973), for a critical review.
\bibitem{annett}
 J.F. Annett, Adv. Phys. {\bf 39}, 83 (1990).
\bibitem{huse}
 D.A. Huse, M.P.A. Fisher, and D.S. Fisher, Nature {\bf 358}, 553 (1992).
\bibitem{fisher}
 M.P.A. Fisher, G. Grinstein, and S.M. Girvin, Phys. Rev. Lett. {\bf 64}, 587
  (1990).
\bibitem{zwerger}
 M. Drechsler and W. Zwerger, Ann. Physik {\bf 1}, 15 (1992); W. Zwerger, in
the
 Proceedings on a Path Integral Conference, May 1992, Tutzing, Germany,
 edited by A. Inamata (to be published by World Scientific, Singapore).
\bibitem{stoof}
 H.T.C. Stoof, Phys. Rev. {\bf B47}, 7979 (1993).
\bibitem{feynman}
 R.P. Feynman, R.B. Leighton, and M. Sands, {\it The Feynman Lectures on
       Physics}, Vol. III, Addison-Wesley, 1965.
\bibitem{greiter}
 M. Greiter, F. Wilczek, and E. Witten, Mod. Phys. Lett. {\bf B3}, 903 (1989).
\bibitem{peng}
 H. Peng and K. Wang, Can. J. Phys. {\bf 69}, 1399 (1991).
\bibitem{ao}
 P. Ao, J. Low Temp. Phys. {\bf 89}, 543 (1992).
\bibitem{dorsey}
 A.T. Dorsey, Phys. Rev. {\bf B46}, 8376 (1992).
\bibitem{yang}
  C.N. Yang, Rev. Mod. Phys. {\bf 34}, 694 (1962).
\bibitem{dirac}
  P.A.M. Dirac, Proc.\ Cambridge Phil.\ Soc.\ {\bf 26}, 376 (1930).
\bibitem{volkov}
 A.F. Volkov and S.M. Kogan, Zh. Eksp. Teor. Fiz. {\bf 65}, 2038 (1973).
\bibitem{penrose}
 O. Penrose, Phil. Mag. {\bf 42}, 1373 (1951).
\bibitem{ufg}
 T. Usui, Prog. Theor. Phys. {\bf 32}, 190 (1964);
 H. Fr\"ohlich, in {\it Problems of Theoretical Physics}, Nauka, Moscow, 1969;
 Z.M. Galasiewicz, {\it Superconductivity and Quantum Fluids}, Pergamon,
 Oxford, 1970.
\bibitem{taylor}
 A.W.B. Taylor, J. Phys. {\bf C3}, L52 (1970).
\bibitem{schiff}
 L.I. Schiff, {\it Quantum Mechanics}, third edition, McGraw-Hill, New York,
 1968.
\bibitem{berry}
 M.V. Berry, in {\it Geometric Phases in Physics}, edited by A. Shapere and F.
 Wilczek, World Scientific, Singapore, 1989.
\bibitem{thouless}
 P. Ao and D.J. Thouless, Phys. Phys. Lett. {\bf 70}, 2158 (1993).
\bibitem{gross}
 E.P. Gross, in {\it Physics of Many-Particle Systems: Methods and Problems},
  Vol. 1, Edited by E. Meeron, Gordon and Breach, New York, 1967.
\end{thebibliography}
\end{document}